\documentclass[onecolumn,sort&compress,numbers]{els-mrw} % For numbered references Style 

\usepackage{amsmath,amssymb,amsfonts,amsthm,makeidx,graphicx}
\usepackage{txfonts}
\usepackage{helvet}
\usepackage{subcaption}

\definecolor{redp1}{rgb}{0.8, 0, 0}
\definecolor{green1}{rgb}{0, 0.5, 0.3}

\begin{document}

%%%%%%%%%%%%%%%%%%%%%%%%%%%%%%%%%%%%%%%%%%%%%%%%%%%%%%%%%%%%%%%%
%% the following items are mandatory: 
%% - title
%% - author names
%% - affiliation details
%% - abstract
%% - keywords

%% Precise, concise, and informative description of the focus of this work. Avoid abbreviations and formulae in the title
\chapter{Long-lived particles: theory and experimental probes}\label{chap1}

%% All author names and affiliations, and email address for corresponding author
\author[1]{Laura Jeanty}%
\author[2]{Brian Shuve}%

\address[1]{\orgname{University of Oregon}, \orgdiv{Institute for Fundamental Science, Department of Physics}, \orgaddress{1371 E 13th Ave., Eugene, OR 97403, USA}}
\address[2]{\orgname{Harvey Mudd College}, \orgdiv{Department of Physics}, \orgaddress{301 Platt Blvd., Claremont, CA 91711, USA}}

\articletag{Chapter Article tagline: update of previous edition, reprint.}

\maketitle

%%%%%%%%%%%%%%%%%%%%%%%%%%%%%%%%%%%%%%%%%%%%%%%%%%%%%%%%%%%%%%%%
%% the following item is mandatory: 
%% 100-150 word summary of the chapter
\begin{abstract}[Abstract]
Long-lived particles (LLPs) are particles that are stable or that live long enough for their decays to be experimentally distinguishable in time or position from their production point. We provide an overview of the phenomenology and experimental signatures of LLPs, focusing on LLPs at the Large Hadron Collider (LHC). We explain what determines a particle's lifetime and we show that LLPs are ubiquitous both within the Standard Model and beyond. We survey the methods used to experimentally detect and characterize particles at collider-based experiments, and discuss how searches for LLPs present both experimental challenges and exciting new possibilities for detection. Finally, we situate LHC searches for LLPs within the broader experimental landscape with a brief overview of searches for LLPs at lower-energy experiments and a discussion of astrophysical and cosmological probes offering complementary insight into the physics of LLPs beyond the Standard Model.

\end{abstract}

%% 5-10 words that embody the key topics in the chapter. What terms would someone put into a search engine if they were looking for a chapter like this?
\begin{keywords}
 	LLPs \sep long-lived particle\sep searches for new particles \sep collider physics \sep Large Hadron Collider \sep physics beyond the Standard Model \sep experimental particle physics
\end{keywords}

%%%%%%%%%%%%%%%%%%%%%%%%%%%%%%%%%%%%%%%%%%%%%%%%%%%%%%%%%%%%%%%%
%% the following item is optional: 
%% - System of abbreviations/terms/symbols used in the specific field of study/community. List and define
\begin{glossary}[Nomenclature]
	\begin{tabular}{@{}lp{34pc}@{}}
		BSM & Beyond the Standard Model of particle physics\\
            EM & Electromagnetic \\
            DM & Dark Matter\\
            LHC & Large Hadron Collider\\
            LLP & Long-lived Particle\\
            MET & Missing Transverse Momentum \\
            SM & Standard Model of particle physics \\
            $\tau$  & A particle's lifetime in its reference frame\\
            $c$ & The speed of light \\
            $\beta = \frac{v}{c}$ & The  speed of a particle as a fraction of $c$ \\
            $\gamma = \sqrt{\frac{1}{1-\beta^2}}$ & A particle's relativistic boost 
	\end{tabular}
\end{glossary}

%%%%%%%%%%%%%%%%%%%%%%%%%%%%%%%%%%%%%%%%%%%%%%%%%%%%%%%%%%%%%%%%
%% the following item is mandatory: 
%% List of the key points and topics a reader can expect to learn from this chapter 
\section*{Objectives}
\begin{itemize}
	\item Define particle lifetime, establish physical properties that affect a particle's lifetime, and assess for a given experiment whether a particle counts as ``long lived''
	\item Survey some common motivations for long-lived particles (LLPs) in theories beyond the Standard Model
	\item Summarize experimental methods for particle identification and characterization at high-energy colliders, demonstrating that LLP signatures are underexplored with significant opportunities and challenges in searches for new LLPs
\end{itemize}

%%%%%%%%%%%%%%%%%%%%%%%%%%%%%%%%%%%%%%%%%%%%%%%%%%%%%%%%%%%%%%%%
%% the following items are mandatory: 
%% - Section: Introduction 
%% - further sections
%% - Section: Conclusion
\section{Introduction}\label{intro}

Most of the elementary particles present in the Universe today are ancient. They survived the inferno following the Big Bang, the formation of galaxies and stars, and the evolution of life. The list of these known, stable particles is relatively short:~protons (themselves made of up quarks, down quarks, and gluons), electrons, neutrinos, and photons. While neutrons are unstable, they can be bound up in stable nuclei giving rise to different isotopes of a given atom. Throw in some dark matter, and you have most of the Universe as it exists today.

But long, long ago, the Universe was a more interesting place, filled in roughly equal measure with six types of quarks, six types of leptons, their antiparticles, all the mediators of the electroweak and strong forces, and more. Most of these particles are unstable and decayed very quickly after the Big Bang to the stable set. Whether or not a particle produced in the Big Bang is still around today depends on its lifetime -- the average time it takes to decay in its own rest frame. Lifetime is a fundamental aspect of a particle's identity.

The Standard Model (SM) consists of particles with a staggering array of lifetimes. The range of lifetimes spans from stable down to $10^{-25}$~s, with a healthy sampling of the total range. Unstable elementary particles are created continuously in interactions that occur in the depths of supernova,  in the halo of black holes, in the collisions between high-energy cosmic rays and our atmosphere. Moreover, the existence of these unstable particles indirectly determines the interactions and behaviors even of the familiar, stable particles, through the presence of their corresponding fields. Understanding the physics of these elementary particles, unstable or not, is the central quest of particle physics -- and it is intimately intertwined with particle lifetime. 

Today, unstable elementary particles are also created in high-energy collisions at particle colliders such as the Large Hadron Collider (LHC). There are compelling reasons to suspect that new particles exist beyond the SM, and that these particles  would also come with a large  range of possible lifetimes. Our experiments must therefore be ready to catch evidence of particles across the full lifetime spectrum. On the short end of the spectrum, many particles decay so quickly that they are never directly registered by the detector, while others are stable and fly entirely through the detector. Some particles decay partway through the detector, producing sprays of decay products like fireworks.  

A particle that survives long enough that its production and decay can be measurably resolved by an experiment is called a {\bf long-lived particle (LLP)}. LLPs are the subject of this chapter. We introduce the fundamentals of lifetime, including the factors that determine a particle's lifetime. We theoretically motivate LLPs, both within the SM and in theories with new particles. We then survey methods for experimentally measuring the properties of LLPs, and talk about how we can discover new LLPs. While we briefly discuss LLPs in the context of a broad range of experiments, we focus on the properties of and searches for LLPs at the LHC because the lessons we can learn about LLPs for that experiment can be readily extended to other experiments. Throughout, we should remember that what counts as an LLP depends on the context of a particular experiment since a short-lived particle in one experiment might look like an LLP to another, depending on the experiment's size and energy scale.

%%%%%%%%%%%%%%%%%%%%%%%%%%%%%%%%%%%%%%%%%%%%%%%%%%%%%%%%%%%%%%%%
%% in the following we showcase possible style elements
%%
%%
\section{Particle Lifetime}\label{sec1}

\subsection{What is lifetime?}\label{sec1:subsec1}

Fundamental particles are curious little creatures. One of their essential properties is that all particles of a certain type are indistinguishable: all muons are identical to all other muons. They don't have an internal clock; there are no young muons and no old muons. Therefore, a fundamental particle's lifetime differs significantly from our intuition based on biological lifetimes. One moment a muon exists, the next moment: poof! there's instead a neutrino, an antineutrino, and an electron. Since muons have no internal memory, this moment cannot depend on when the muon came into existence; there is a fixed decay probability at every single moment. The probability of a given particle decaying within a fixed time interval is described by a Poisson distribution. In turn, the probability $P$ that a particle has not decayed by time $t$, measured in the lab frame and given that it was around at time $t=0$, is given by an exponential distribution:
\begin{equation}
    P(t) = e^{-\frac{t}{\gamma\tau}}
\end{equation}
where $\tau$ is the lifetime of the particle in its rest frame, and $\gamma= \frac{1}{\sqrt{1-\frac{v^2}{c^2}}}$ is the Lorentz boost of the particle, where $c$ is the speed of light. One of the essential properties of the exponential distribution is that its shape is scale invariant once you fix $\tau$; this is a necessary property for the distribution that describes the collective decay probability of particles, since the particles have no memory of what came before.

Note that nuclear physicists often use a half-life $\tau_{1/2}$, the time by which a particle has a 50\% probability of having decayed, whereas particle physicists use the mean lifetime $\tau$. The two are related by $\tau = \frac{\tau_{1/2}}{\ln2}$. After a time $t=\tau$ in its rest frame, a particle has a $\frac{1}{e}\approx 37\%$ chance of not having decayed. 

\begin{figure}[h]
	\centering
	\includegraphics[width=15cm]{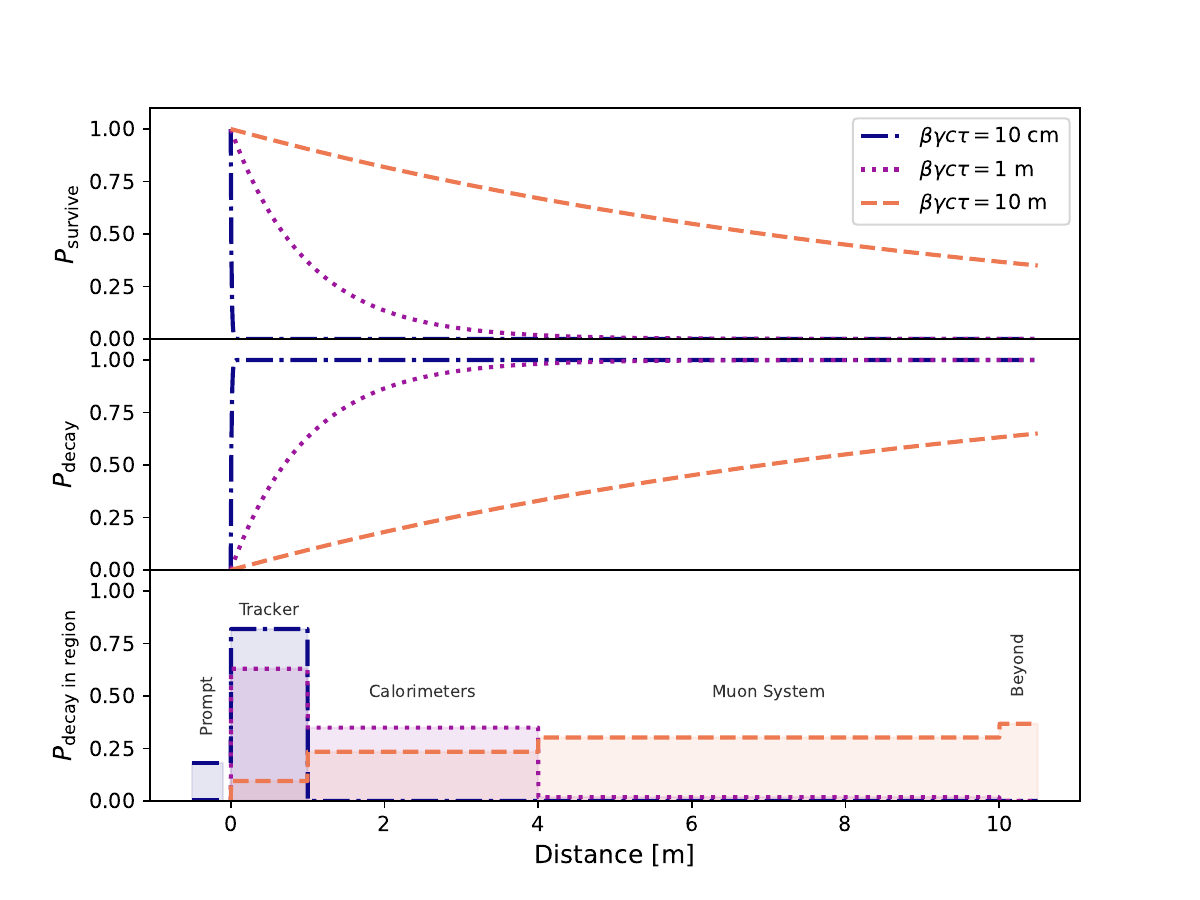}
	\caption{A particle's mean decay position, $d$, is determined by its rest-frame lifetime $\tau$ and speed through $d=\beta\gamma c\tau$. Top panel: The probability for a particle to not have decayed by position $d$, for three different values of $\beta\gamma c\tau$. Middle: the probability for a particle to have decayed by position $d$. Bottom: the probability that the particle decays within a given detector volume, using the regions defined by Figure~\ref{fig:detector}, where the "Prompt" region is defined by $d < 2$ mm.}
	\label{fig:exponential}
\end{figure}

In the lab frame, a particle moving at a speed $v$ with lifetime $\tau$ will have a 63\% chance of decaying before traveling a distance $d = \beta \gamma c \tau$, where $\beta \equiv \frac{v}{c}$ and $d$ is the distance in the direction of motion. The probability of survival, decay, and the chance of decay within a given detector region are illustrated in Figure~\ref{fig:exponential} for three different values of $\beta\gamma  c\tau$. Although we may approximate the distance that a high-energy particle travels as $d \approx c\tau$, this only works for a narrow range around $\beta\gamma \approx 1$ and, in any case, a particle will only make it that far 37\% of the time. Since $\beta\gamma = 1$ when $\beta = \sqrt{\frac{1}{2}} \approx 0.7$, particles moving more slowly than $0.7c$ will not travel as far as $c\tau$ on average. In contrast, highly boosted particles can have $\beta\gamma \gg 1$, and therefore can travel significantly farther than $c\tau$. Recall that $\beta\gamma = \frac{p}{mc}$, where $p$ is the magnitude of the momentum of the particle, which gives a simple way to estimate the boost of a particle.

 Although only 37\% of the particles will survive to a time $t=\tau$ in their rest frame, the mean decay time for a collection of particles with the same lifetime will still be $\tau$ due to the long tail of the distribution. Recall that for an exponential distribution, both the mean and the standard deviation are equal to $\tau$, which makes it convenient to estimate $\tau$ for an exponential distribution without having to perform a fit. Another property to note from the exponential is that the distribution peaks at $t=0$, which indicates that -- regardless of lifetime -- a single particle has the largest probability of decaying around $t=0$ (and hence $d=0$) for any fixed time (or distance) interval.

\subsection{What determines lifetime?}

While lifetime is a fundamental property of a particle, it differs from some other properties like charge and spin. The latter are intrinsic to a particle and, in part, define it. A particle's lifetime, however, depends on the masses, charges, spins, and couplings of other particles with which it can interact.  Fermi's golden rule tells us that the decay rate depends on the square of the quantum mechanical amplitude for the transition, as well as the multiplicity of available final states in the decay. More quantitatively, the width of a particle, defined to be $\Gamma \equiv \frac{\hbar}{\tau}$, is
\begin{equation}
\Gamma \propto \frac{1}{2m}\int |\mathcal{M}|^2 d\Pi,\label{eq:golden_rule}
\end{equation}
where $m$ is the mass of the decaying particle, $\mathcal{M}$ is the matrix element or amplitude for the particle's decay to a particular set of other particles, and the integral is over the momenta, or phase space, of all final particles ($d\Pi$ is a product with a factor of $\frac{d^3p}{2E}$ for every final-state particle, as well as overall energy- and momentum-conserving delta functions). If a particle can decay via multiple channels, then $\Gamma = \Sigma_i\Gamma_i$, where $\Gamma_i$ is the partial width in each channel $i$. Often the lifetime is dominated by the decay into one or two channels. 

As the width of a particle is inversely proportional to its lifetime, we can measure the lifetime of short-lived particles by measuring the width of an observable, such as cross section, as a function of collision energy or mass. Long-lived particles have widths that are narrower than experimental resolution, and the lifetime can instead be measured by fitting an exponential distribution to decay position or time. At current experimental capabilities, there is an intermediate regime in which we can not directly measure a particle's lifetime as it is too short to observe that it decays at a later time than it is produced but too long to give a measurable width. The Higgs boson is currently in this intermediate regime; instead, its lifetime can be indirectly measured via observables such as the ratio of off-shell to on-shell Higgs bosons, which is sensitive to the width.

\subsection{What are long-lived particles and why are they long lived?}\label{sec:what_are_LLPs}

As a qualitative descriptor, ``long lived'' depends on a reference scale. A muon's lifetime of $2.2\,\,\mu$s might not seem long by the human scale, but it's positively ancient in the particle world where objects are regularly zipping by at or close to the speed of light. Practically, a particle is usually considered long-lived in a given experiment if the experiment has the resolution to observe any evidence of a non-zero lifetime. Otherwise, it's called ``short-lived'' or, more often, ``promptly-decaying,'' and its decay products are ``prompt.'' The ability to detect a non-zero lifetime depends on the time and position resolution of the detector as well as the relativistic boost of the particle in question. At the LHC, particles with lifetimes greater than about 1 ps are generally considered long-lived. 

Based on Eq.~\eqref{eq:golden_rule}, a particle can acquire a long lifetime if either the amplitude $\mathcal{M}$ or the integral over phase space $d\Pi$ is small for the leading (or only) decay mode. We can categorize the mechanisms by which one of these factors is small in three ways:

\begin{enumerate}
    \item The decay proceeds through a heavy intermediary particle with mass $M$, where $M \gg m$. This heavy scale shows up in the denominator of a propagator in calculating $\mathcal{M}$ and suppresses the matrix element;
    \item A small coupling appears in $\mathcal{M}$;
    \item The integral over phase space $d\Pi$ is small. This occurs when the mass of the decaying particle $m$ is only fractionally larger than the sum of the masses of its decay products or when the mass of the decaying particle itself is small. In these cases, only a small set of momenta (and hence available final states) are accessible in the decay.
\end{enumerate}

The Standard Model is full of particles that have long lifetimes due to the above mechanisms. We'll list a few favorite examples here. The muon gets its long lifetime from the heavy $W$ boson through which it must decay ($M_W \gg m_{\mu}$), which effectively makes the weak interaction ``weak'' at the muon mass scale. Kaons (bound states of one strange quark and one up or down quark) have much longer lifetimes than the $D$ mesons (bound states of one charm quark and one lighter anti-quark) because of the quirks of the CKM matrix, which gives $|V_{cs}| > |V_{us}|$. In other words, strange quarks talk more strongly to $c$ quarks than to $u$ quarks, giving a larger effective coupling in $\mathcal{M}$. The smallness of $|V_{ub}|$ and $|V_{cb}|$ are also responsible for the relatively long lifetimes of $b$ hadrons compared to $D$ mesons. Finally, the neutron acquires its tremendous lifetime of 15 minutes through a small integral over $d\Pi$; the neutron mass is so close to the sum of the masses of the proton, electron, and neutrino to which it decays ($\Delta m < 1$ MeV) that there is very limited kinematic phase space to sample in its decay.

Note that a fundamental particle's lifetime can differ in a bound state. First, one must take into account hadronic effects of both the initial and final state when considering hadronic decays. Additionally, bound-state considerations for the decay products can also affect the energy configurations of the decay products, including binding-energy effects and considerations like the Pauli exclusion principle. The impact of the bound-state effect is nowhere clearer than with neutrons; while $\tau = 15$ minutes seems luxurious in terms of detecting a lifetime, it would be rather detrimental to the universe as we know it if 63\% of the neutrons in all atomic nuclei disintegrated every 15 minutes. 

\begin{figure}[h]
	\centering
	\includegraphics[width=15cm]{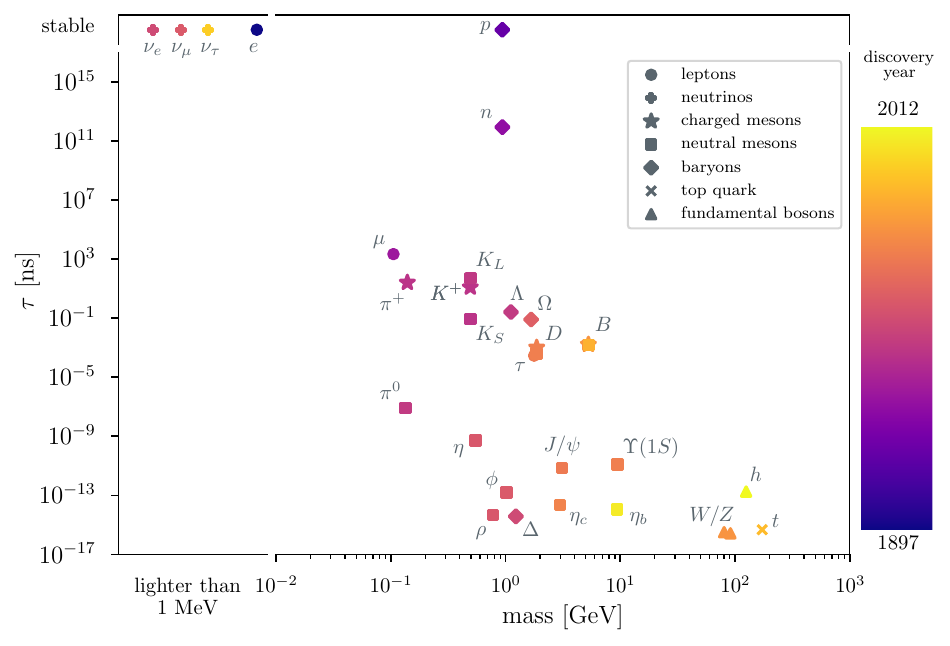}
	\caption{A selection of Standard Model particles in the mass-lifetime plane. Different markers indicate different categories of particles, while the shading indicates the year the particle was discovered. While the order in which particles are discovered depends on many factors, there is an apparent trend where the earliest particles to be discovered within a given species have relatively long lifetimes. This figure is inspired by earlier versions from Refs.~\cite{Alimena:2019zri,Lee:2018pag}.}
	\label{fig:SM_lifetimes}
\end{figure}

Given that the lightest particle of each type has few (or no) decay modes accessible, the lightest particle of each sector tends to have a long lifetime -- and indeed is often stable. The electron, the up and down quarks, the photon, the pion, the neutron, the proton... they are all long lived. Add to this the fact that lighter particles are easier to produce, and the result has been that the first particle of a new type to be discovered has almost always been long lived. This is shown in Figure~\ref{fig:SM_lifetimes}, which illustrates the complex relationship between a particle's mass, lifetime, type, and year of discovery for a selection of SM particles intended to highlight the major discoveries of new fundamental particles. Of course, the relationship between these properties depends on many other factors, but there nonetheless remains a strong association between when a new type of particle was discovered (lepton, boson, baryon, first generation meson, second generation meson, etc.) and its lifetime.

\section{Theories that predict new LLPs}\label{sec:theory}

There are many reasons to believe that new particles exist beyond the ones that we've discovered so far. And just as many SM particles are LLPs, we expect that at least some of these new particles will be LLPs too. In fact, Figure~\ref{fig:SM_lifetimes} suggests that if we extrapolate from the existing experimental record, we're likely to discover a new sector first through its long-lived states. 

In the sections below, we'll briefly summarize a few leading motivations for new sectors beyond the SM, and show how LLPs naturally fit into these frameworks. These examples are only a small subset of the theories that predict new particles and forces with LLPs; indeed, the mechanisms that give SM long lifetimes, enumerated in Section~\ref{sec:what_are_LLPs}, easily and naturally arise in many extensions of the SM \cite{Curtin:2018mvb,Lee:2018pag,Alimena:2019zri}. 

\subsection{Dark matter}
When we look at our galaxy or out into the broader universe, we come to a startling conclusion: the amount of matter that we measure through its gravitational pull on nearby stars and gas is much larger than can be accounted for by the light we collect in telescopes! The missing matter, which is invisible to every detection method we have devised so far, is called ``dark matter'' (DM). There is over five times more of it than regular ``visible matter,'' which nowadays is mostly made of protons, neutrons, and electrons. Amazingly, in every moment of the Universe's history that we can see -- from the first light in the Universe from before atoms formed (the cosmic microwave background) down to the present day -- we see evidence for DM. The broad consensus among physicists and astronomers is that DM is made up of at least one type of new, stable particle.

The gravitational evidence suggests that there exists at least one new LLP -- DM itself! -- and, potentially, many others. Whatever particle(s) make up DM should have lifetimes of at least the age of the Universe, since otherwise they would have decayed away a long time ago. 

Let's imagine for a moment that DM, whatever it is, bears some resemblance to the lightest stable particles in the Standard Model: protons, neutrons, electrons, and neutrinos. These particles that persist to the present are but a fraction of all the SM particles that once existed. Muons, taus, pions:~they are all LLPs that live a little while but then decay away. It seems likely that DM is the same way. Maybe there are dark muons, dark taus, dark pions, with lifetimes much shorter than DM itself but still LLPs. If this is true, we say that dark matter is part of a ``dark sector,'' a whole collection of new particles and forces of which we are as-yet unaware. These shorter-lived dark LLP particles might be easier to detect than the truly invisible DM because, when they decay, they could give off visible signals in particle colliders. 

{\bf Example: inelastic dark matter.} Inelastic DM is a dark matter model where there are two dark states:~the lighter one, $\chi_1$, is the true DM particle, while the heavier one, $\chi_2$, can decay into the lighter one through processes like $\chi_2\to \chi_1+e^+e^-$, or  $\chi_2\to \chi_1+q\bar{q}$. DM could be produced at the LHC through related processes like $q\bar{q}\to\chi_2\bar\chi_1$. The model is called ``inelastic'' because a quark-antiquark pair creates a $\chi_2\bar\chi_1$ pair, but not $\chi_1\bar\chi_1$ or $\chi_2\bar\chi_2$ pairs.

Inelastic DM generically occurs in situations where DM matter feels its own dark electromagnetic force mediated by a dark photon. Occasionally this dark photon talks to regular electrically charged particles too like electrons and quarks. Unlike our photon, however, the dark photon has a mass; otherwise, its interactions could lead to disasters like our Sun cooling too quickly by giving off dark-photon radiation. Generically, inelastic DM models produce two dark states that are very close in mass, $\Delta M \equiv M_{\chi_2}-M_{\chi_1}\ll M_{\chi_1}$. Because the energy released in $\chi_2$ decays is very small, its lifetime can be very long ($c\tau\propto \Delta M^{-5})$ according to point \#3 in Sec.~\ref{sec:what_are_LLPs}, so $\chi_2$ is often an LLP \cite{Bai:2011jg,Izaguirre:2015zva}.

\subsection{Sterile neutrinos} \label{sec:sterilenu}
Neutrinos were originally thought to be massless; the existence of neutrino masses was a surprise confirmed at experiments that detected \emph{neutrino oscillations}. If neutrinos were massless, the only thing that would distinguish one neutrino from another would be the flavor of lepton with which it is produced in $W$ or meson decays (for example, if $W^+\to\mu^+\nu$, then the neutrino produced is a $\nu_\mu$). If neutrinos have mass, however, then the neutrino produced in association with a muon is a superposition of different mass eigenstates. Under time evolution, the result is a non-stationary state that can oscillate from one flavor of neutrino to another. 

SM matter particles get their masses by interacting with the Higgs field, and the SM neutrinos are likely no exception. Massless spin-$1/2$ particles come in two types, depending on how they spin relative to their direction of motion:~left-handed (LH), and right-handed (RH). To give particles a mass, the Higgs field unites  a LH particle and a RH particle. The SM neutrinos are observed to all be LH, so the existence of neutrino masses suggests the existence of three new particles, the RH neutrinos. In the Sterile neutrino model, when the Higgs field turned on in the early Universe, these LH and RH states got shuffled and mixed up, with the result that today there are \emph{two} types of neutrinos:~the SM neutrinos, which feel the weak force and have very tiny masses ($\sim0.1$ eV), and the sterile neutrinos, which do not feel the weak force and are typically much heavier. In this scenario, the SM and sterile neutrinos are each their own antiparticles, which can have interesting implications for cosmology and neutrino experiments. For example, because sterile neutrinos are their own antiparticles, they could decay into SM leptons \emph{or} SM antileptons, and an imbalance in these decays could explain why there's more matter than antimatter in the Universe.

The SM neutrinos are stable LLPs, and since they only experience the weak force, they escape most experiments undetected. As the name implies, sterile neutrinos typically have tiny couplings to SM leptons, and so they can also be LLPs, albeit meta-stable ones. Because the sterile neutrinos are heavier than SM neutrinos, they can decay into SM neutrinos and other SM leptons, unlike neutrinos which have nothing lighter to decay into. 

{\bf Example:~low-scale seesaw neutrinos.} In this model, there exist two or more right-handed neutrinos (RHNs) that couple to the left-handed neutrinos (LHNs). The SM neutrino masses are generated by the mechanism described above. The model is called a seesaw because, for a fixed coupling between RHNs and LHNs, the SM neutrinos are light when the sterile neutrinos are heavy. In the most famous version of the seesaw model, the sterile neutrinos are extremely heavy (with masses $\sim10^{14}$ times the proton mass), and this gives a natural explanation for why the SM neutrinos are so light. Alternatively, sterile neutrinos with smaller couplings could have masses within reach of current colliders (masses $\sim1-100$ GeV).

For simplicity, we consider the case of a single sterile neutrino, $N$. Because $N$ is mostly made of RHN states but has a tiny component of LHN, it does feel the weak force but much more feebly than SM neutrinos. For every SM process that produces a SM neutrino, such as $W^+\to\ell^+\nu$, there is a tiny probability that a sterile neutrino is produced instead, $W^+\to\ell^+ N$. In turn, this sterile neutrino could decay back to the SM neutrino, $N\to W^+\ell^-$. But, hang on:~if $N$ is lighter than the $W^+$ boson and produced in $W^+$ decays, it wouldn't have enough energy to decay back into a $W^+$! $N$ would instead decay through a \emph{virtual} $W^+$ boson; the lifetime of $N$ would naturally be long according to point \#2 in Sec.~\ref{sec:what_are_LLPs}. So, $N$ is typically an LLP \cite{Atre:2009rg}!

\subsection{Supersymmetry}

Imagine if every particle in the SM had a fraternal twin:~similar, but not the same. In this scenario, each particle is paired with a so-called superpartner that has the same charges under the SM forces, but different quantum properties. When a SM particle is a fermion, like an electron, the corresponding superpartner is a boson (in this case, a ``selectron'').

This might sound far-fetched, but it's actually a natural consequence of a hypothesized spacetime symmetry called \emph{supersymmetry}, or SUSY for short \cite{Martin:1997ns}.  There are many good reasons to believe that SUSY is a fundamental symmetry of spacetime. It solves a major issue in the SM called the hierarchy problem, which comes from the fact that quantum effects tend to make the Higgs boson much heavier than we observe it to be. SUSY also naturally comes with a DM candidate in the form of the lightest superparticle, if that particle is stable and electrically neutral. 

While the lightest SUSY particle could be DM and therefore neutral (and hard to directly produce at colliders), heavier SUSY particles would carry electroweak and strong charges and would be abundantly produced. In many cases, the next-to-lightest SUSY particle would be an LLP.

SUSY is a major experimental target for several reasons. First, it is very well motivated from a theoretical point of view. Second, the huge number of new SUSY particles means that there are a lot of different ways to look for it. Since we don't know the masses of any of these particles, we don't know which ones we expect to find first at the LHC. The landscape of SUSY signatures is valuable not just for searching for SUSY, but for systematically covering many signatures that appear in numerous varieties of new physics models. The landscape of SUSY signatures includes many possible LLPs, including split SUSY \cite{Arkani-Hamed:2004ymt,Giudice:2004tc}. 

{\bf Example:~gravitino DM SUSY.} In some SUSY scenarios, the lightest SUSY particle is the gravitino ($\tilde G$), which is the superpartner of the graviton\footnote{The masses of the SUSY particles depend on something called the SUSY breaking scale, which determines how much heavier the superpartners are compared to SM particles. Gravitino DM naturally arises in a scenario where the superpartner masses are determined through a mechanism called gauge mediation \cite{Meade:2008wd,Knapen:2016exe}.}. The coupling between the gravitino and other SUSY particles is suppressed by inverse powers of a heavy mass scale, and so the decay of the next-to-lightest SUSY partner to the gravitino would occur very slowly; it is often an LLP according to point \#1 in Sec.~\ref{sec:what_are_LLPs}. For gravitino DM SUSY, the next-to-lightest SUSY particle is often the superpartner of the tau lepton (the ``stau'', $\tilde\tau$), which would decay as $\tilde\tau\to\tau 
\tilde{G}$. Note that in some scenarios the stau lifetime could be so long that it would make it all the way through the detector before decaying.

\section{Experimental Detection of LLPs}
\subsection{Everything we see is an LLP}

Experimental design is motivated by physics goals and tempered by constraints, including technological, financial, logistical, and other considerations. Radiation poses one of the most significant constraints on a collider detector, determining how close a detection element can be to the collision point: the zillions of particles cumulatively produced in colliders cause damage to the detector, one interaction at a time. At the LHC, the detectors are, at closest, a few cm away from the collision point. Only particles that live long enough to cross at least some part of the detector, $\Delta t\sim10^{-10}$ s in the lab frame, can be directly measured. In that sense, every particle we directly detect interacting with our detector is an LLP. 

The  list of SM particles that fit the bill is short. These particles include protons, neutrons, electrons, muons, neutrinos, photons, and long-lived hadrons, including pions and kaons. When a particle is ejected from the collision, it is the detector's job to determine what type of particle it is, and to measure its energy and momentum four-vectors as carefully as possible. The first task is called particle identification, and the second is kinematic reconstruction. Modern detectors are designed to do these two jobs very well. In this section, we describe how experimentalists directly detect particles produced in a collider.

Collider detectors have many layers like a leek, or like a parfait. The inner layers of the detector, collectively called a ``tracker,'' are designed to precisely measure the helical trajectory of a charged particle moving through a magnetic field, using as little material as possible to minimize interactions that change the particle's trajectory. Modern trackers are built from materials like semiconductors or gases that ionize when a charged particle passes through. Each detecting element is divided into pixels or strips, and by recording which pixels or strips have ionization signals above threshold, we can trace the particle's path. Only charged particles ionize, so only charged particles, including protons, $\pi^+$, electrons, and muons, leave tracks in the tracker.

The second subsystem of the detector consists of calorimeters. These are designed with dense materials to initiate electromagnetic and hadronic interactions with incident particles, causing them to slow down, deposit their kinetic energy into the detector, and eventually stop. Electromagnetic braking interactions (Bremsstrahlung) with the detector occur on a shorter length scale than hadronic interactions of colorless hadrons, so an inner EM calorimeter captures mainly electromagnetic showers while an outer hadronic calorimeter captures the energy from hadronic showers. By measuring the ratio of energies deposited in each calorimeter as well as the shower shape and size, as well as the shower shape and size, we can distinguish whether the particle was, for example, more likely a photon or a neutron.

Muons lose some energy due to ionization in the calorimeters, but as Bremsstrahlung is inversely proportional to the incident particle mass squared, an electron loses roughly 40,000 times more energy due to Bremstrahlung than a muon of the same energy. As muons don't interact hadronically, they pass through the calorimeters unstopped. To measure their momenta more precisely and to identify their tracks as muons, the calorimeters are surrounded by another, enormous tracker, generally called a muon system.

\begin{figure}[h]
	\centering
	\includegraphics[width=12cm]{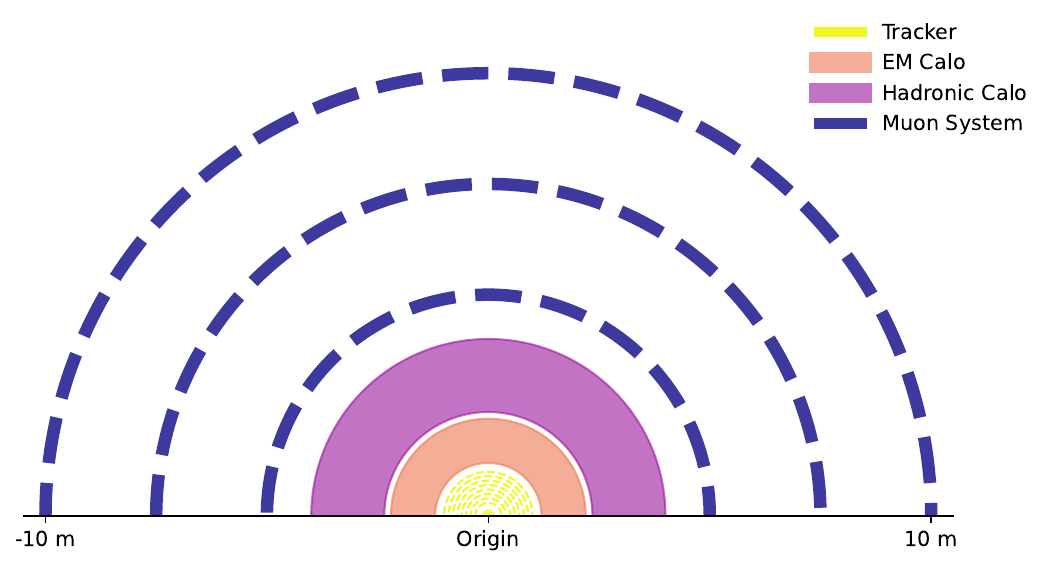}
	\caption{A transverse cross-section view of a simplified detector at the LHC, where the beam runs perpendicular to the page. Only the top half of the detector is shown. The inner tracker consists of silicon pixel and strip layers surrounded by an electromagnetic calorimeter, which is in turn surrounded by a hadronic calorimeter. Outside the calorimeters, a gaseous tracking detector serves as a muon system.}
	\label{fig:detector}
\end{figure}

A cartoon schematic of a collider detector at the LHC is shown in Figure~\ref{fig:detector} in transverse view, with the beam perpendicular to the page. This cartoon is a simplification of the actual detectors, both in terms of geometry, which is less homogeneous and symmetric than the figure shows, and in terms of elements displayed. Only the active sensing elements are shown; not displayed are the support structures, cooling systems, magnets, power and data cables, etc. The inner tracker is a simplified view of the actual detectors, which additionally include, for example, a layer of small gas tubes (the TRT in ATLAS) or a Cherenkov particle identification layer (the RICH detector in LHCb).

\subsection{Indirect particle detection for Standard Model particles}

With a limited range of particles that we can directly detect at colliders, we need additional methods to indirectly detect the remaining particles in the SM, including $Z$ bosons, Higgs bosons, top quarks, and neutrinos. The first three of these particles all rapidly decay into SM LLPs. There are two primary ways that decayed particles are indirectly detected: through kinematics, and through vertexing. 

Consider, for example, that a Higgs boson can decay into two photons ($H\to\gamma\gamma$). Once it has decayed, these photons at face value seem similar to photons produced in any other process at the LHC, such as in the decay of neutral pions or radiated from charged particles. However, Higgs bosons leave a subtle imprint in properties of the photons they produce. Precise measurement of the kinematic properties of the two photons allows us to reconstruct the di-photon invariant mass, which should be equal to the Higgs boson mass if the photons came from a Higgs boson. We can then infer the existence of the Higgs boson statistically by looking for an excess at a particular mass, as opposed to photon pairs produced by other SM processes which will have an invariant mass spread over a wide range of values. There are other ways that kinematics are useful:~for example, the typical lab-frame energies and angles of the photons, along with other SM particles produced in association with the photons, can differ for Higgs decays compared to other sources of photons. 

The second method is through vertexing. Since we can trace the paths of charged particles through the inner detector, we can see whether these paths originate along the beam line where collisions occur. We say that SM particles produced directly in the initial collision are ``prompt'' or ``primary'' particles. In the case of Higgs bosons, which decay in less than $10^{-20}$ s, the point of origin of the decay products is indistinguishable from the initial collision point, and so we also call these decay products prompt.

There are, however, instances where a parent particle travels a finite distance before decaying. One of the most important examples of these at the LHC is the bottom quark. Bottom quarks at the LHC have typical boost factors $\gamma\sim10$ and lifetimes $\sim10^{-12}$ s. Consequently, they travel a distance of about 3 mm before decaying. When the bottom quark decays and the tracks of the decay products are reconstructed,, it is possible to tell that they originate from a few mm away from the collision point rather than the collision point itself. We call such particles ``displaced,'' and it is one of the clearest signs that one particle has been produced in the decay of another. When we see multiple charged particles originating from a point far from the beamline, we can reconstruct a displaced vertex. This property of $b$-quark decays allows us to separate their decays from the decays of other quarks with a high degree of accuracy.

Neutrinos only interact via the weak force, and are not slowed down or stopped by any detector layer. They are invisible from an experimental point of view. To infer their existence, detectors which cover a symmetric and significant fraction of the total solid angle measure a quantity called ''Missing (Transverse) Momentum," or MET.\footnote{Confusingly, MET is also commonly referred to as ``missing transverse energy,'' even though energy is a scalar quantity without an associated direction.}.'' In a hadron collider, the total transverse momentum of the parton--parton collision is close to zero, which must be true before and after the collision. A vector sum of all measured momenta after the collision yields a magnitude and azimuthal angle, $\phi$, for the momentum that is missing in the transverse direction, indicating an invisible particle (or several) carried away both energy and momentum. In a lepton collider, the total initial momentum is known in all three spatial dimensions, and total missing momentum can be inferred from the total sum of measured momentum.

\section{Finding new LLPs beyond the SM}

\subsection{What determines search strategy?}

As discussed in Sec.~\ref{sec:theory}, there could be many new LLPs out there that we haven't discovered yet.
The best strategy to search for each LLP varies, and depends on the LLP properties. The most important properties that determine search strategy include:
\begin{itemize}
    \item {\bf Mass and Production mechanism:}~if LLPs are produced in the decays of heavy particles (with masses $\gtrsim\mathrm{TeV}$), then they or their decay products are very energetic and typically leave more distinctive signals in the detector. If, instead, LLPs are produced in the decays of the Higgs boson, $Z$ boson, or similar particles (with masses $\sim100\,\,\mathrm{GeV}$), or if the LLPs are low in mass and directly produced, then the LLPs will have small energies compared to typical LHC energy scales and could be harder to distinguish from SM processes. Different search strategies are employed for the two different cases.

    \noindent There is another factor to consider. In some cases, an LLP could be produced in association with another particle. For example, the low-scale seesaw neutrino we introduced in Sec.~\ref{sec:sterilenu}, $N$, could be produced in $W^\pm$ decays like $W^\pm\to e^\pm N$. In this case, sterile neutrinos would always be accompanied by an electron or positron. This $e^\pm$ could be tagged and reconstructed to reduce backgrounds and improve prospects of finding the LLP in this channel.
    
    \item {\bf Lifetime:}~LLPs decay according to an exponential distribution with lab-frame lifetime $\gamma\tau$. The decay therefore occurs at a mean distance $\gamma\beta c\tau$, although the most likely decay distance is still at the origin. Therefore, at distances $d\gg \gamma\beta c\tau$, an LLP has an exponentially tiny survival probability and it can only be seen indirectly by analyzing its decay products. If $d$ is large compared to the detector size, we expect many LLPs to survive all the way through the detector, with some fraction of them decaying inside. If $d$ is small compared to the detector size, then all of the LLPs will decay, and their signatures depend on where they decay.

    \item {\bf Charges under SM forces:}~how the LLP appears in the detector depends on whether and how it is charged under SM forces, such as the EM or strong force. When LLPs carry EM or strong charges, they directly interact with the detector as they travel outward from the collision point. If the LLP carries no SM charges or is charged under only the weak force, it won't leave a direct trace in the detector.

    \item {\bf Decay products:}~Assuming the LLP decays inside the detector, it can be indirectly reconstructed from its decay products. How an LLP decays affects how we look for it. If the LLP decays into quarks or gluons, its decay products can be more often mimicked by SM processes due to the huge rate of strong-interaction processes at the LHC. Decays into multiple charged leptons, on the other hand, would be more distinct, motivating a different search strategy and improved sensitivity to leptonically decaying LLPs.
    
    \item {\bf Boost:}~a particle's boost is determined by the relativistic factor, $\gamma = \frac{E}{mc^2}$. How does this affect an LLP's properties? First, a particle with $\gamma$ close to one $(v< c)$ will move slowly, unlike SM particles produced at the LHC with moderate or higher momenta ($p \gtrsim 1$)~GeV/c. The LLP could therefore be distinguished from SM processes by the fact that it's moving slowly. Second, conservation of momentum in the LLP rest frame necessitates the outgoing particles to have balanced momenta in different directions once it decays. If the LLP is relativistic in the lab frame, these decay products become collimated with a small angle relative to the parent LLP's direction. Suppose the LLP goes in the $x$-direction in the lab frame. Then all particles emitted in the hemisphere centered on the $x$-axis in the LLP rest frame will be compressed into a cone of half-angle  $\tan^{-1}\left(\frac{1}{\beta \gamma}\right)$ as seen by observers in the lab frame. 
    
\end{itemize}

\subsection{Direct searches for new LLPs}

The classic experiments of early particle physics often involved observing a new particle directly interacting with the detector; for example, the discovery of the positron and muon were both detections of a particle leaving traces in a cloud chamber that were inconsistent with known particles. The tracks of the positron and muon could be distinguished from those of electrons and protons by using the measured curvature, ionization signature, and range to extract the charge and mass of the incident particles. 

\begin{figure}
\centering

\begin{subfigure}{0.33\textwidth}
\centering
\includegraphics[width=1.0\linewidth]{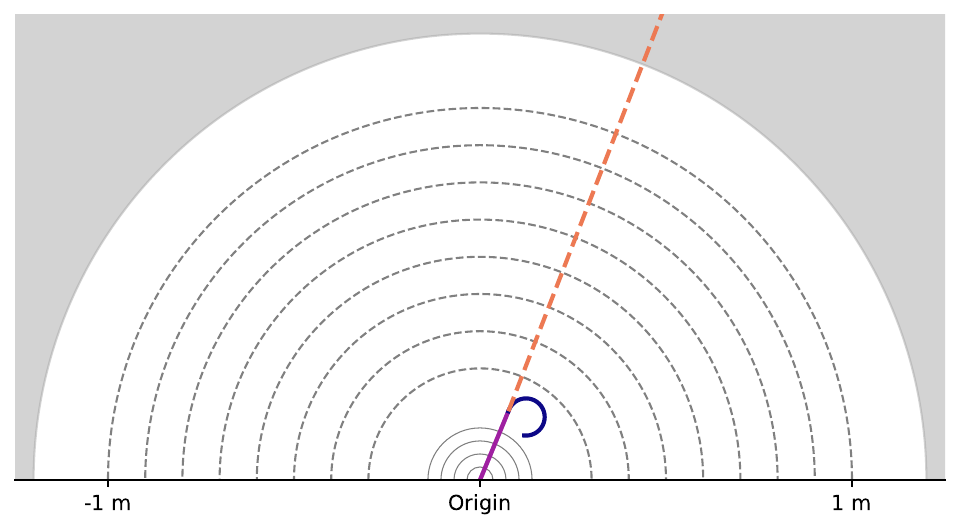}
\caption{A disappearing track signature}
\label{fig:disappearing_track}
\end{subfigure}
\begin{subfigure}{0.33\textwidth}
\centering
\includegraphics[width=1.0\linewidth]{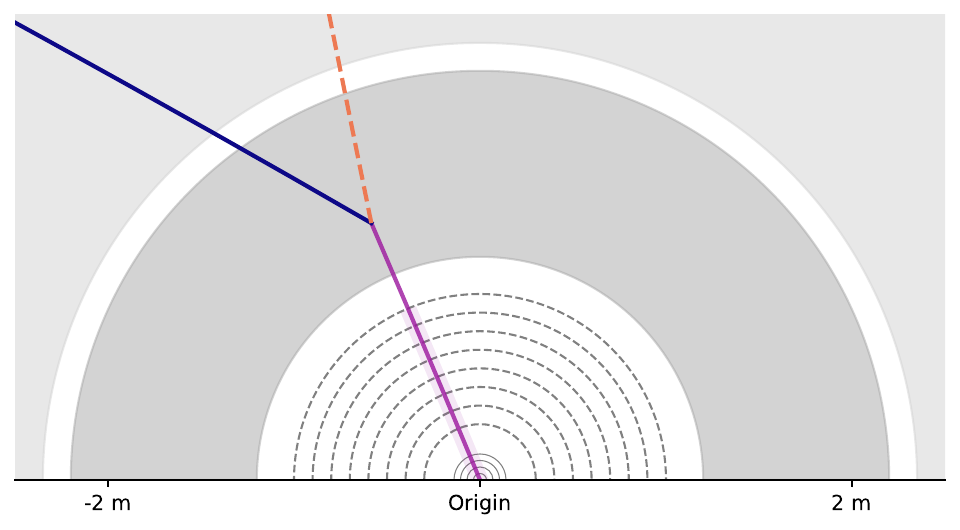}
\caption{An ionization signature}
\label{fig:dedx}
\end{subfigure}
\begin{subfigure}{0.33\textwidth}
\centering
\includegraphics[width=1.0\linewidth]{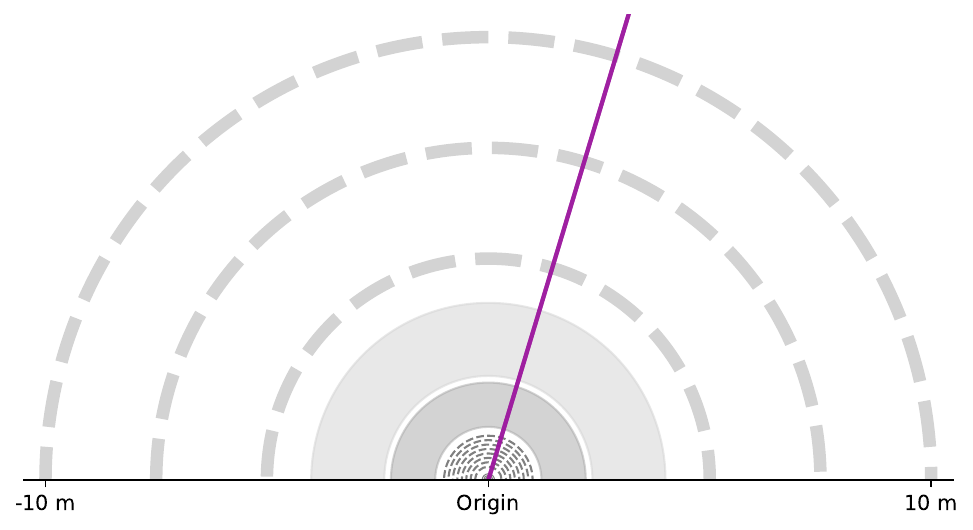}
\caption{A track from a charged LLP}
\label{fig:smp}
\end{subfigure}

\caption{Several examples of direct detection signatures from electrically charged LLPs. In the first panel, an LLP decays inside the tracker into a charged particle with very low momentum and a neutral particle, creating a disappearing track signature. In the middle panel, a heavy, charged LLP leaves more energy due to ionization energy loss in the tracker than a minimum ionizing particle with comparable momentum. The LLP decays in the calorimeter, but the details of its decay are not relevant to the $dE/dx$ signature. In the right panel, a charged LLP does not decay inside the detector and therefore leaves a track throughout the whole detector, which can be identified as originating from a heavy BSM particle due to $dE/dx$ or Time-of-Flight measurements in any of the detector subsystems.}
\label{fig:direct_detection}
\end{figure}

Direct searches for new LLPs today follow the same path and indeed use modern versions of the same tools and techniques to look for particles interacting in the detector in ways that are inconsistent with any SM particle. Direct searches do not usually depend (much) on how a new particle might decay, and therefore have broad sensitivity to a wide swath of potential new LLPs, as long as they interact with the detector and have a lifetime long enough to get to (and through!) the relevant detector element. At the LHC, $\gamma\tau \ge \mathcal{O}(0.1)$ ns is a convenient threshold beyond which direct detection can be reasonably efficient. We show some typical LLP direct detection signals in the toy detector in Fig.~\ref{fig:direct_detection}.

Most direct searches for new LLPs look for particles that are electrically (or occasionally magnetically) charged. To distinguish the track of a new LLP from an SM LLP, a search will focus on properties of tracks that are sensitive to a particle's mass (assuming the new particle is heavy) or charge (if the new particle has $q \neq 1$ or any magnetic charge), kinematics, or decay position. It is possible to build models with a new LLP that is electrically neutral but charged under the strong force; here, hadronic interactions with detector material could make direct detection possible, as well, as occurs with the neutron.

As charged particles undergo helical motion in a magnetic field, tracks are fit with a 5-dimensional helix. A common parametrization of the helix returns the charge-to-momentum ratio of a track ($\frac{q}{p}$), two angles of the track ($\phi$ and $\theta$), and the longitudinal and transverse impact parameters ($z_0$ and $d_0$). $z_0$ is a measure of how close the extrapolated trajectory is to the collision point along the direction of the beam, while $d_0$ is a measure of how close the extrapolated trajectory is to the collision point in the transverse plane. These 5 track parameters are sufficient for the vast majority of collider physics measurements and searches. However, to determine the mass of a particle from its track, we need more information and we must measure additional quantities beyond the particle's trajectory. If a detector element can precisely record the time at which a particle interacts or the charge that it deposits as it traverses the element, we can measure $\beta$, or equivalently, $\beta\gamma$. Since $\beta\gamma = \frac{p}{mc}$, measuring $\beta\gamma$ and $p$ allows us to directly extract $m$. As the LHC doesn't produce SM LLPs with lifetimes $>$ 1 ns which are heavier than the proton at 1 GeV$/c^2$ (or 2 or 3 GeV$/c^2$ if one considers the rare deuteron and even rarer triton), looking for a track with $m \gtrsim $ a few GeV is a robust way to look for new LLPs. 

We can measure $\beta$ from a track by measuring the Time-of-Flight (ToF), which is the time it takes for a particle to reach a given detector layer, assuming that the particle was produced at the origin of the detector at $t=0$. Comparing the measured ToF for the particle to the ToF for a particle traveling to the same detector position at $v=c$ allows us to measure $\beta$. To make a useful measurement of $\beta$, the time resolution of the detector must be significantly better than the time differences we want to measure. The latter are determined by distance to the detector position and the expected $\beta$ for the particles of interest. To claim that a particle with a given $\beta$ is traveling more slowly than $c$ with a confidence level of $n\cdot\sigma$, a detector at a distance $d$ from the origin must have timing resolution $\sigma_t$ such that:
\begin{equation}
    \Delta t = \frac{d}{c}\left(\frac{1}{\beta}-1\right) \ge n \cdot\sigma_t.
\end{equation}
For example, if we want to measure that a particle traveling at $\beta = 0.8$ is traveling more slowly than $c$ with a confidence of $3\sigma$ based on a detector measurement at $d=10$~m, we find that the detector must have a timing resolution of 3 ns or better.

Certain detectors can also measure the ionization energy loss, $dE/dx$, along a track. Specifically, the average energy loss $dE$ per unit length $dx$ for thick detectors is given by the Bethe-Bloch equation:
\begin{equation}
-\frac{dE}{dx} = 0.3071\, q^2 \frac{Z}{A}\frac{1}{\beta^2}\left( \ln{\frac{2m_ec^2(\beta\gamma)^2}{I}-\beta^2+\frac{\delta}{2}} \right)\, {\frac{\texttt{MeV}}{\texttt{g/cm$^2$}}}
\end{equation}
where $q$ is the incident particle charge in units of the electron charge, $Z$ and $A$ are the atomic number and weight of the detector material, $m_e$ is the electron mass, $I$ is the mean excitation energy of the material, and $\delta$ is a density parameter that reduces the ionization for very relativistic particles~\cite{Grupen:2008zz}. The Bethe-Bloch equation is normalized to the density of the material in units of $\texttt{g/cm$^3$}$ -- to get $\frac{\texttt{MeV}}{\texttt{cm}}$, multiply $dE/dx$ by the density of your detector material. Note that for thin silicon sensors or small gaseous tubes such as those used in modern trackers, the average energy loss is not a well-defined quantity or experimentally useful, and instead one should use the most probable value of energy loss~\cite{BICHSEL2006154}. For a given material, and assuming that the incident particle has $q=1$, $dE/dx$ depends only on the incident particle's $\beta\gamma$. Therefore, measuring $dE/dx$ gives a direct measurement of $\beta\gamma$, and combining this with $p$ one can directly extract a particle's $m$. 

If one is looking for LLPs with $q \ne 1$, then the standard calculations of $p$ from the track curvature are incorrect, but the track curvature and $dE/dx$ can still be useful discriminants. For magnetically charged objects, Dirac predicted that the fundamental magnetic charge is $q_m = Ng_Dq_e$, where $N$ is an integer, $q_e$ is the charge of the electron, and the Dirac charge $g_D = \frac{1}{2\alpha}$. This predicts that for $N=1$, a magnetic monopole would ionize like a particle with $q = 68.5$ -- more than 4,000 times more ionization than for an electric monopole! 

In addition to non-standard tracks due to non-standard charge, charged LLPs in certain models give rise to tracks with strange trajectories due to decays or interactions in flight, including tracks that change direction suddenly (kinked tracks) or tracks that disappear (disappearing tracks)~\cite{Fukuda_2018}. Kinked or disappearing track signatures arise from charged LLPs that decay to one very low-momentum charged particle whose momentum is too low to be reconstructed (or selected) and a neutral decay product which does not interact with the detector. In models in which the LLPs are charged under non-SM forces, tracks could even have non-helical trajectories~\cite{Kang_2009}. Additionally, in experiments with Cherenkov light detectors, it is possible to tag a slow-moving (heavy) LLP by the absence of Cherenkov light~\cite{LHCb:2015ujr}.

\subsection{Indirect searches for new LLPs}

If an LLP is electrically neutral and has no strong charge, it will not interact with the detector as it travels. If it is stable or its lifetime is long enough that the vast majority of LLPs produced at a collider travel through the entire detector before decaying, the only signature of the LLP is missing momentum or MET. Searches at colliders for many models of dark matter involve looking for evidence of a new, neutral, long-lived LLP via observables derived largely from MET. \footnote{Note that in the case of both neutrinos and DM candidates, there is a large program of completely independent experiments (called "direct detection" in the latter case \cite{Billard:2021uyg}) which look for evidence of the neutral LLP interacting with a detector in a single interaction via the debris or recoil from the interaction. This is a different categorization of direct detection than the one we introduced above, since even in a DM direct-detection experiment, the DM interactions are too weak to reconstruct anything like the track of the DM-candidate as it moves through the detector. The best one can hope for is a single interaction.} 

For LLPs with a lifetime small enough that a substantial fraction decay within the detector volume, we can search for the LLP via "indirect detection" if it is possible to detect its decay products. At the LHC, $\mathcal{O}(0.1)$~ps $\le\tau \le \mathcal{O}(30)$~ns is a rough lifetime range in which indirect detection is possible, although the precise sensitivity depends on the $\beta\gamma$ of the produced LLP and which detector system is used to detect the decay products. In general, indirect detection relies on identifying one or more of the decay products as displaced from the primary collision in either space or time (or both). The optimal search strategy depends on the LLP mass, decay mode, and decay position. If an LLP decay includes one visible particle and one neutral, weakly or non-interacting particle, we can search for a displaced object. If the LLP is pair-produced and both LLPs decay in the detector, we can look for two displaced objects. 

If the LLP decay includes a photon and occurs before the calorimeter, the displaced photon can be identified as "non-prompt" by using the spatial resolution of the electromagnetic calorimeter to tag that the photon does not point back to the interaction point, and/or by using the timing resolution of the calorimeter to tag that the photon arrives later than a photon produced in the primary interaction. Electrons can be tagged by the calorimeter as non-prompt similarly to a photon. Energy deposits from an LLP that decays inside a calorimeter can produce unconventional shower shapes relative to prompt jets, since the LLP-seeded jet starts at some point inside the calorimeter rather than at the collision point. Depending on the radial segmentation of the calorimeters and the LLP decay position, the shower shape can also be used to tag a displaced photon, electron, or jet. If an LLP is electrically neutral and has no strong charge, it will not interact with the detector as it travels. 

\begin{figure}
\centering

\begin{subfigure}{0.33\textwidth}
\centering
\includegraphics[width=1.0\linewidth]{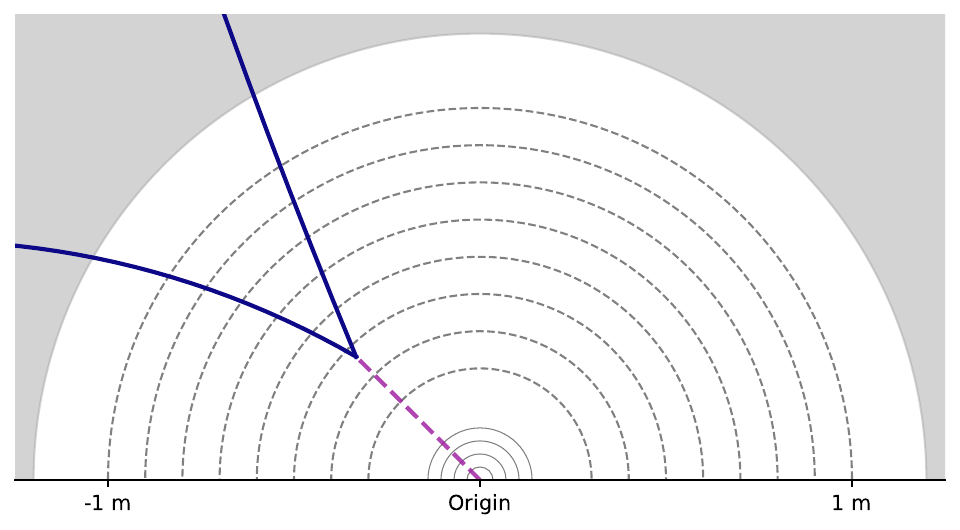}
\caption{A displaced-vertex signature}
\label{fig:basicDV}
\end{subfigure}
\begin{subfigure}{0.33\textwidth}
\centering
\includegraphics[width=1.0\linewidth]{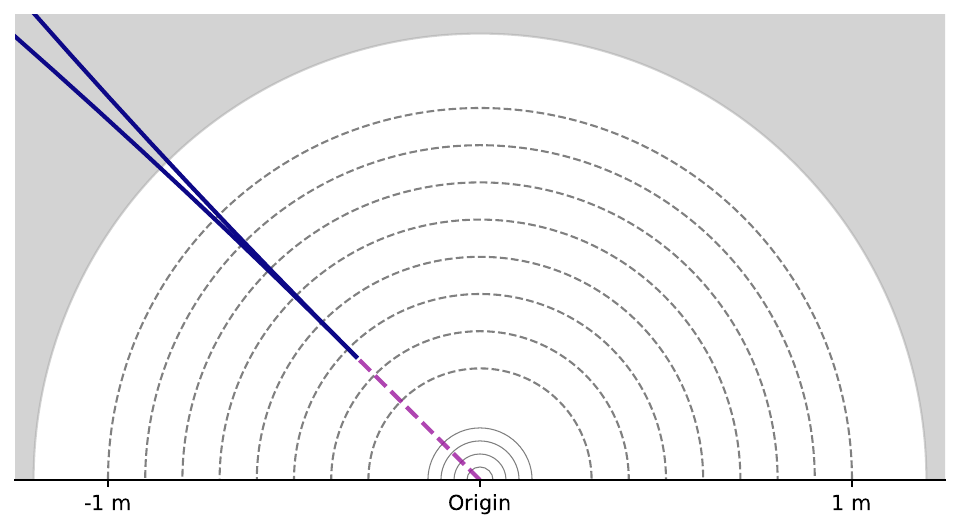}
\caption{A boosted displaced-vertex signature}
\label{fig:boostedDV}
\end{subfigure}
\begin{subfigure}{0.33\textwidth}
\centering
\includegraphics[width=1.0\linewidth]{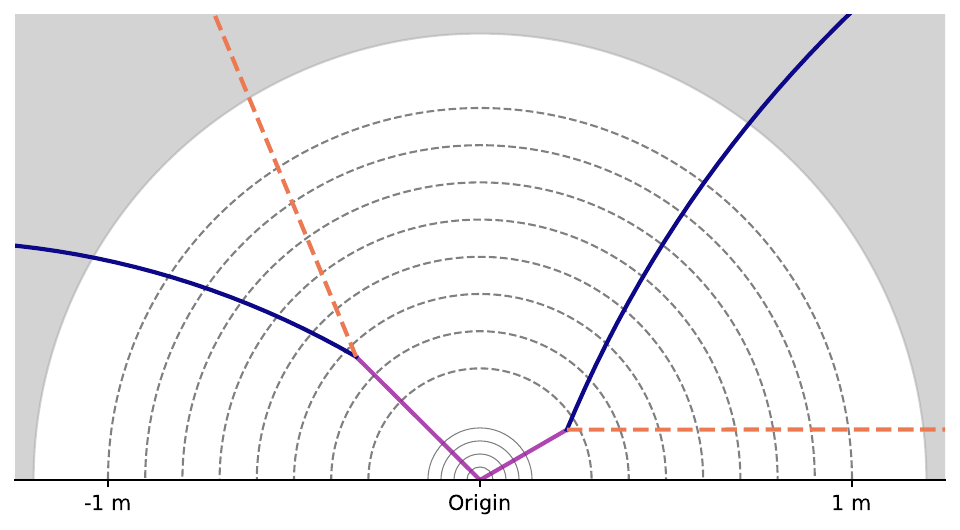}
\caption{A signature of two displaced tracks}
\label{fig:boostedDV}
\end{subfigure}

\caption{Several examples of signatures involving displaced objects in the tracker. In the first panel, a heavy, neutral LLP decays to a pair of charged particles inside the tracking detector layers. Tracks from the charged decay products can be reconstructed as displaced based off both their origin as well as their impact parameters, and a displaced vertex can be reconstructed. In the middle panel, a boosted LLP produces a pair of displaced tracks. The tracks are displaced in origin, and a displaced vertex can be reconstructed. However, as the decay products inherit the parent LLP's boost, they are collimated, and it may not be possible to use the track impact parameters to identify that the tracks are displaced as they will point back to the origin. In the right panel, pair-production of charged LLPs, each of which decays to a charged and neutral decay product, produces a signature with two displaced tracks which do not form a displaced vertex.}
\label{fig:DVs}
\end{figure}

For LLP decays that include a charged particle, this particle can also produce a track, if it traverses a significant fraction of the tracker or muon spectrometer. There are at least three ways to identify a track as displaced: 1) the track originates after the first layer (or better yet several layers) of the inner tracker, in which case we identify a well-reconstructed track with no hits in the first few layers; 2) the track has impact parameters $d_0$ and potentially $z_0$ which are inconsistent with the primary vertex of the main collision point in the event; and/or, 3) if the primary LLP is moving slowly or if its decay produces a longer path length to a given detector element, a charged decay product may arrive at a tracker detector late, and timing can help identify the track as displaced. Note that for light LLPs, their decay products will not necessarily arrive later than a SM counterpart, so timing is not always useful to identify displaced tracks. Similarly, a track can originate at a large radius in the detector but still have small $d_0$ and $z_0$, for example in the case of a boosted LLP which decays to two light decay products; these decay products will still tend to point back to the collision point. 

For LLPs that decay to at least two charged particles, a "displaced vertex" can be reconstructed from two or more displaced tracks by reconstructing the common origin of the tracks. Several indirect-detection signatures involving displaced tracks and vertices are shown in Fig.~\ref{fig:DVs}. Since a reconstructed vertex requires a common origin of at least two reconstructed objects, an LLP that decays to a single charged particle (along with other neutral particles) produces a displaced track but not a displaced vertex. Similarly, an LLP which decays inside a calorimeter, even if to multiple charged particles, may form a displaced jet but not a displaced vertex. Depending on the LLP lifetime, a displaced vertex may also be reconstructible for an LLP that decays to two photons, if the photons can be resolved as two distinct objects in the calorimeter, or if the photons convert to $e^+e^-$ pairs through interacting with material in the detector and the momenta of the $e^+e^-$ pairs point back to a position that is displaced from the collision point. Once a displaced vertex is identified, its invariant mass can be calculated from all tracks originating from the vertex ($\sqrt{(E^{\rm tot}_{\rm vtx})^2-(\vec{p}^{\rm tot}_{\rm vtx})^2 }   $). In the case that an LLP decays to fully reconstructible objects, this should reconstruct the LLP mass. In the case that the decay also includes neutrals (pions, neutrinos, DM candidates, etc.) then the reconstructed invariant mass is smaller than the LLP mass, but often much higher than SM backgrounds.

\begin{figure}
\centering

\begin{subfigure}{0.33\textwidth}
\centering
\includegraphics[width=1.0\linewidth]{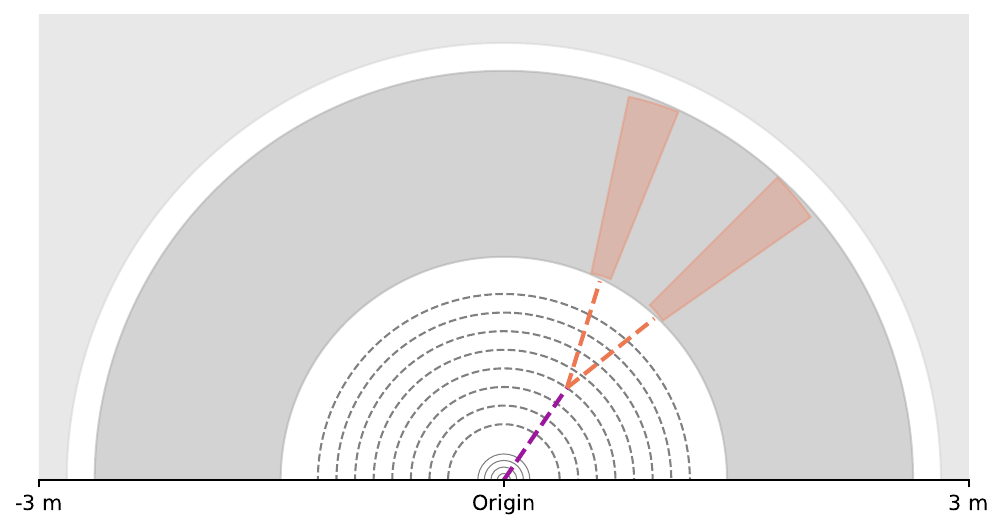}
\caption{A non-prompt photon signature}
\label{fig:basicDV}
\end{subfigure}
\begin{subfigure}{0.33\textwidth}
\centering
\includegraphics[width=1.0\linewidth]{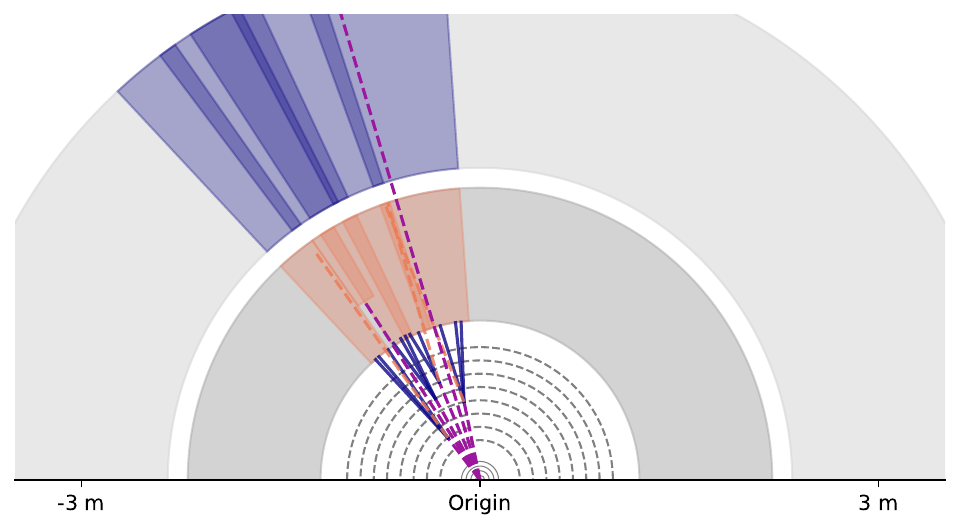}
\caption{An emerging jet signature}
\label{fig:boostedDV}
\end{subfigure}
\begin{subfigure}{0.33\textwidth}
\centering
\includegraphics[width=1.0\linewidth]{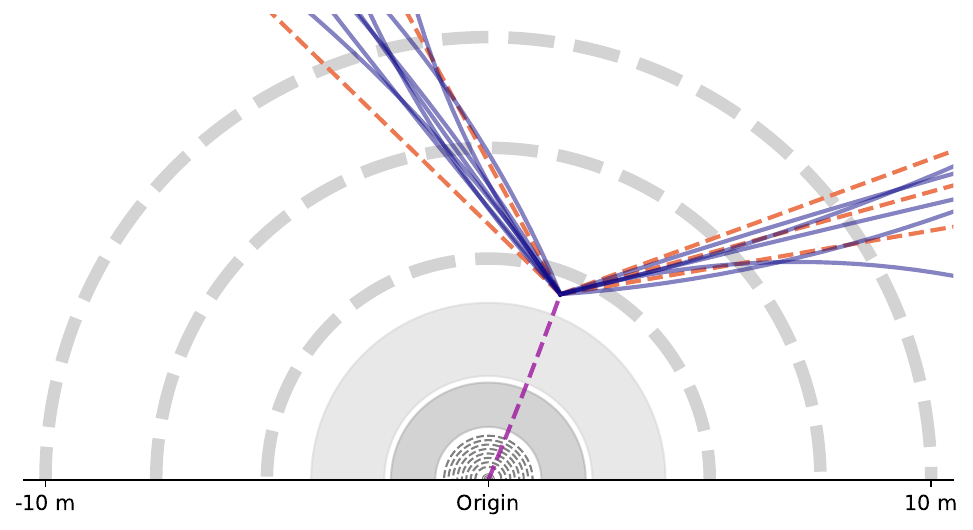}
\caption{Displaced jets in the Muon System}
\label{fig:boostedDV}
\end{subfigure}
\caption{Several examples of signatures involving displaced objects in the calorimeters and muon system. In the first panel, a heavy, neutral LLP decays to a pair of displaced photons, which show up as showers in the electromagnetic calorimeter. They can be identified as coming from an LLP based on timing or pointing signatures. In the middle panel, a jet of dark hadrons produces an emerging jet signature, in which individual dark hadrons decay into SM hadrons at different positions. In the right panel, a neutral LLP decays into two hadronic jets at the beginning of the muon system.}
\label{fig:indirect_2}
\end{figure}

Some models of new physics could produce striking final states that include more than one displaced signature. For example, some theories predict a large number of LLPs that are produced in a jet in a manner similar to SM hadrons \cite{Strassler:2006im}. The result is that there could exist one or more collimated cones of multiple displaced objects as each LLP decays back to SM particles at a different position. This "emerging jet" signature could be reconstructed as multiple displaced vertices \cite{Schwaller:2015gea}, or as a jet with unusual shower shape, depending on the decay positions. Figure~\ref{fig:indirect_2} illustrates several displaced signatures, including emerging jets. If an LLP that is charged under the EM or strong force is produced nearly at rest, the energy it loses due to ionization or hadronic interactions can be sufficient to cause it to stop within the detector. If the lifetime of such an LLP is long enough, it can sit within the detector until long after the collision that produced it, producing a striking decay signature much later, even in an otherwise empty event.

\subsection{Challenges in experimental searches for new LLPs}\label{sec:challenges}

By construction, all the fundamental particles we haven't yet discovered are difficult to find for one reason or another; otherwise, we would have found them already! For BSM LLPs, the experimental challenges are numerous. The first challenge is simply the vast phase space of potential signatures, which is amplified by the unknown LLP lifetime. The unknown lifetime cannot simply be scanned as an additional parameter; like mass or coupling strength; different search strategies are needed for different regimes. Since usually the mass and coupling strength are additionally unknown, the LLP lifetime axis leads to a multiplicity of signatures, each of which often requires a dedicated search. Moreover, the same signature in different lifetime regimes will require information from different sub-detectors, which must be reconstructed, processed, calibrated, and analyzed separately. 

Beyond the sheer vastness of signatures, additional challenges in searching for BSM LLPs mainly stem from the fact that most experiments today are optimized to identify SM LLPs from prompt BSM decays. They are not optimized to search for BSM LLPs. Therefore, designing an LLP search often requires developing custom data collection methods and storage formats, reconstruction algorithms, calibration techniques, and background estimation methods. In a modern particle physics experiment, each of these steps can require significant investment of time and effort -- and/or provide significant limitations on analysis sensitivity.

The first and irreversible step of data filtering is the trigger. The collision rate at the LHC and other collider experiments is too high to store and reconstruct data from every collision. The trigger identifies interesting collisions to store for future study, while data from all other collisions are discarded. The trigger typically selects SM objects -- such as electrons, muons, photons, and jets -- that originate from the collision point. As both direct and indirect BSM LLP signatures differ from SM and prompt BSM signatures, the default triggers that an experiment develops can be highly inefficient for BSM LLPs. 

One exception is the MET trigger, which often serves as a catchall with some trigger efficiency for LLP events, especially for heavy LLPs. The first-level MET trigger relies on a vectorial sum of calorimeter information; as long as the new LLP is heavy ($\gtrsim 100$~ GeV) and does not decay entirely to SM hadrons, electrons, and/or photons before the calorimeter, there is usually some trigger efficiency possible, of order 10\% or higher. The energy imbalance results from the fact that the emission of either strong or electromagnetic initial-state radiation (ISR) will be reconstructed in the calorimeter, but the full LLP energy will not be reconstructed. For low-mass LLPs which don't come along with significant ISR, or for all-hadronic final states, the MET trigger can be highly inefficient and other trigger pathways, including muon triggers and/or jet triggers, might have some efficiency instead. 

In many cases, developing a custom trigger for a given LLP signature will  significantly increase the trigger efficiency. Other strategies, such as data scouting (saving limited object information for a larger number of events that otherwise would not pass the trigger), data parking (saving raw data to be processed when more resources are available), or improving real-time reconstruction to identify LLP signatures at the trigger level, can be employed to significantly improve the trigger efficiency for LLP searches \cite{LHCb:2023hlw,CMS:2024zhe}.

After the trigger, experiments usually process and significantly reduce the information stored in data formats used for analysis, in order to efficiently use both disk space and computing resources. This can easily throw away information that is valuable for LLP searches including timing and charge information, high granularity calorimeter cluster energy distributions, etc. Therefore, LLP searches often require custom data formats or early data processing, which can be constrained by computing resource considerations. Clever optimizations and schemes must be developed to save or add back information that is uniquely essential for LLP searches.

Once the data has been triggered and the relevant information has been included in an appropriate data format, the LLP signature of interest must be reconstructed and calibrated. Often, custom reconstruction is required, since default reconstruction algorithms typically assume that particles originate from the collision point and will make optimizations that are inefficient for displaced or slow moving particles. These optimizations can include, for example, discarding hits recorded with a timestamp that is inconsistent with a particle moving at $\beta = 1$, or only considering hits within a path that points back to the interaction point, or discarding seed jets that do not have energy deposits in the first layer of the calorimeter. Whether or not custom reconstruction is required, custom calibration is often required for analyses which use nonstandard low-level detector information. For example, if precise timing information is used, small changes in cavern temperatures can affect the signal propagation time in cables, which affects the apparent time measurement by a detector. As another example, radiation damage causes loss of charge collection efficiency in silicon detectors. If this charge information is used to calculate a $dE/dx$ measurement, then the measurement must be calibrated for the effect of radiation damage, which depends on integrated luminosity, temperature profiles, high voltage, annealing periods, etc.

An additional challenge for LLP searches is that the sources of background are often largely instrumental, meaning that the background is not dominated by one or two SM processes with final-state signatures similar to the LLP signature of interest. Rather, backgrounds come from sources including: the tails of reconstructed kinematic distributions, mis-reconstructed objects, and the accidental overlap of unlikely processes (the unlikely processes include many sources, including prompt particle production, cosmic rays, detector noise, interactions of beam particles, etc). These backgrounds are hard to properly simulate, and therefore many LLP searches must design fully data-driven methods of estimating the background.

In addition to the challenges noted above in selecting an LLP signature and estimating the background remaining after the selection, an analysis must also estimate the uncertainties associated with these steps. While this is true for all searches, it often requires dedicated procedures for LLP searches, as the triggers/reconstructed objects/calibrations/background techniques are custom-built. Estimating uncertainties for LLP searches is also complicated by the fact that there is often no similar SM object on which to test procedures in the data.

While the challenges are numerous, the flip side is that LLP signatures are underexplored compared to other searches for new physics at the LHC. This is exciting from two points of view: 1) there is still significant room for discovery of new particles, and 2) there is room for analysis innovation. The extensive and innovative LLP search programs at ATLAS, CMS, and LHCb, along with the dedicated LHC experiments discussed below, continues a long tradition of creative collider-based LLP searches, and will continue into the high luminosity LHC era and beyond.

\section{The Current LLP Experimental Landscape}

\subsection{LHC}
Most of this article is focused on LHC experiments, especially the general-purpose ATLAS and CMS detectors. These detectors are large, sensitive to particles emerging in almost any direction from the collision, and are hermetic (without cracks). This gives them excellent sensitivity to LLPs with lab-frame decay lengths from mm to m and beyond, for both direct and indirect detection. They also record the largest number of collisions of any of the LHC experiments. There has therefore been a sustained and robust LLP search program at both ATLAS and CMS.

LHCb is another major LHC experiment with a significant LLP search program. The LHCb detector differs significantly from ATLAS and CMS, as it is designed primarily to make precision measurements of $b$ quarks. $b$ quarks at high boost ($\beta\gamma \gg 1$) are predominantly produced at a small angle relative to the beam. The LHCb detector therefore consists of trackers and calorimeters targeting the forward region close to the beam, and only on one side. $b$ quarks are SM LLPs with relatively short proper decay lengths ($c\tau\approx0.5$ mm), but their relativistic boost, especially in the forward region, allows LHCb to efficiently distinguish their decays from the collision point. Since LHCb was designed to look for SM LLPs, it is a great environment to look for BSM LLPs, and the large boost of particles typically observed in the detector allows it to probe shorter proper lifetimes than ATLAS and CMS typically do. There are limitations, however, largely stemming from the limited angular coverage of the detector and the fact that it observes a lower rate of collisions compared to the ATLAS and CMS experiments.

\begin{figure}[h]
	\centering
	\includegraphics[width=10cm]{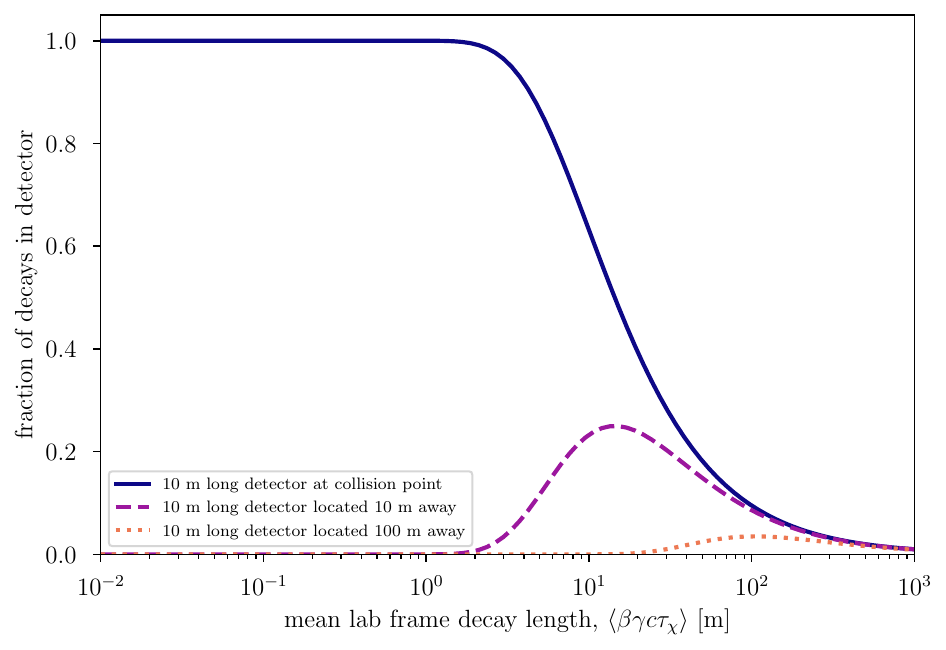}\hspace{1cm}\includegraphics[width=10cm]{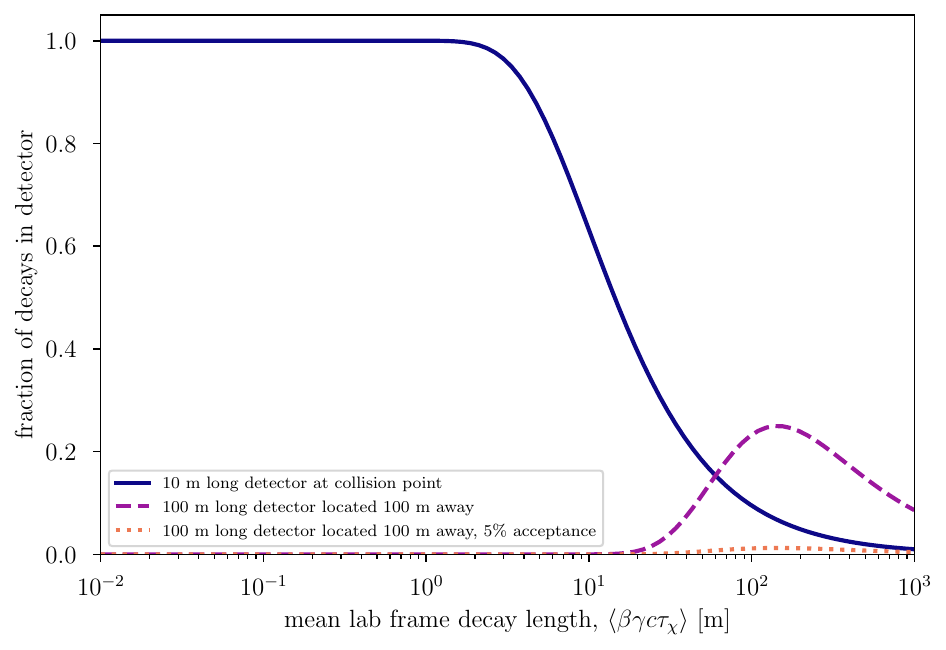}
	\caption{In these plots, we illustrate the fraction of LLPs that decay in a given detector as a function of the mean lab-frame decay length, $\langle\beta\gamma c\tau\rangle$, and the detector geometry. (Top) The detectors are all 10 m in length (perpendicular to the beam) but located different distances from the collision point. We assume that the detectors completely surround the collision point. It is evident that the detector located directly at the collision point has the best performance. (Bottom) Now, we compare a 10 m long detector located at the collision point (solid) to a 100 m long detector located 100 m from the collision point (dashed). The 100 m detector has superior performance at long lifetimes due to its larger size. However, it is unrealistic to expect such a detector to completely surround the collision point. We therefore also show a detector of same length perpendicular to the beam that only covers 5\% of the solid angle from the collision (dotted). This performs at best comparably to the smaller detector at the collision point, and often much worse. }
	\label{fig:acceptance}
\end{figure}

As outlined in Sec.~\ref{sec:challenges}, there are many challenges that limit the sensitivity of the primary LHC experiments to certain LLP models, especially at lower LLP masses and/or when the LLPs decay into SM particles carrying strong charge because of the huge background rate. This has led to many proposed {\bf dedicated detectors}, which are new detectors placed far from LHC collision points that are constructed specifically to look for LLPs (either directly or indirectly). Some of these ideas, such as the FASER \cite{FASER:2022hcn} and milliQan experiments \cite{Ball:2016zrp}, have come to fruition and are already running, while others are still in the proposal stages. The common feature of dedicated detectors is that they are shielded from the intense radiation emanating from the collision point. This reduces backgrounds and allows the dedicated detector to efficiently trigger on and reconstruct LLP signals. The price these detectors pay is that they only span a small solid-angle region of particles traveling out from the collision point. The larger the detector, the better the sensitivity, although there are examples such as the FASER detector (which targets LLPs produced along the direction of the LHC beam) where even modest-sized dedicated detectors have good sensitivity to LLPs.

In Fig.~\ref{fig:acceptance}, we plot the fraction of LLP decays in a given detector configuration as a function of the mean lab-frame decay distance, $\langle\beta\gamma c\tau\rangle$. We find that, for fixed detector size, the optimal position for maximum acceptance is always to have the detector centered at the origin since the origin is the most likely decay position for any exponential distribution. For longer lifetimes, a longer detector has a better chance of capturing LLP decays, but if this large detector only has a small solid-angle coverage of LLPs emerging from the collision point, the gain due to length is largely mitigated by the reduced acceptance. This shows that the main benefit of dedicated detectors far from the LHC collision point is not necessarily that a larger fraction of LLPs decay inside the detector, but rather that the location and shielding suppress backgrounds from the primary collision, allowing more efficient triggering and reconstruction of LLP signals.

\subsection{Electron-Positron Colliders}
While the highest-energy collider operating today is the LHC, there are also electron-positron colliders operating at lower energies. The most important of these for LLP searches are the B factories, which operate at a CM-frame collision energy of $\sqrt s=10.58$ GeV to  produce and study copious numbers of B mesons for the purpose of precisely studying properties of $b$ quarks such as their decays and $CP$ violation (different behaviors for particles and antiparticles). Since $b$ quarks have proper decay distances $c\tau\sim0.5$ mm, the detectors at B factories are specifically designed to be able to spatially resolve their decay positions relative to the collision point. While the design goal of B factories is to do precision studies of SM physics, these experiments could also be producing all sorts of new BSM particles through $e^+e^-$ annihilation, including LLPs, so long as these new particles have masses below about 10 GeV. The only currently operating B factory is Belle II \cite{Belle-II:2018jsg}, although previous B factories BABAR and Belle \cite{BaBar:2014omp} continue to perform new studies with their data.

There are several advantages to LLP searches at B factories compared to the LHC. Because the initial state experiences  the electromagnetic force rather than the strong force, the background rate is lower. The overall collision rate is lower and so triggers can also be more generous in deciding which collisions to record. Because the initial $e^+e^-$ state is, in most cases, completely annihilated in the collision, it is possible to use energy and momentum conservation to fully reconstruct invisible LLPs (including their masses) produced in the collision.

There are a few disadvantages of B factories as well. Their relatively low CM energies mean that they are only sensitive to LLPs lighter than about 10 times the proton mass. This means they are sensitive to BSM LLPs with similar masses and lifetimes to SM LLPs, and consequently many LLP signatures are not as distinctive as at higher-energy colliders. The B-factory detectors are smaller than their counterparts at the LHC, which reduces sensitivity in the long-lifetime limit. Finally, because B factories are designed to precisely study SM particles rather than to discover new BSM particles, the person-power on these experiments is predominantly allocated to studies of SM particle properties, not BSM particle production. Nevertheless, there has been a growing effort to improve B-factory sensitivity to new particles, including LLPs.

There are other, lower energy $e^+e^-$ colliders, such as BES III~\cite{Yuan:2019zfo}, which have CM-frame collision energies up to about 5 GeV. In the past, there were also higher-energy $e^+e^-$ colliders such as LEP, which operated up to 209 GeV. While LEP's integrated luminosity is small compared to the B factories and the LHC, it offers good sensitivity to LLPs with masses below 100 GeV provided they are produced at a sufficiently high rate. It is challenging to go to higher electron energies in a circular accelerator due to beam synchrotron radiation:~proposals for future high-energy lepton colliders include linear accelerators~\cite{Myers:1991ym, Bambade:2019fyw}, very-large-radius circular accelerators~\cite{Benedikt:2928193}, or muon accelerators~\cite{InternationalMuonCollider:2025sys}.

\subsection{Fixed-Target and Beam-Dump Experiments}

It is much easier to rapidly collide high-energy particles by slamming them into a target at rest (fixed-target experiments) rather than colliding them into other high-energy particles moving in the opposite direction. 
The upside of fixed-target experiments is the very high collision rate, which gives them sensitivity to extremely rare processes. The downside of fixed-target experiments is that the CM-frame energy is much lower than a head-on collider:~the CM-frame collision energy is instead $\sqrt{s}=\sqrt{2E_{\rm beam}M_{\rm target}}$, where the typical target is a proton ($M_p\approx1\,\,\mathrm{GeV}/c^2$) inside of a nucleus.

Fixed-target experiments should have a thick-enough target that most incident beam particles undergo scattering, while the target should be thin enough to allow the final-state particles to escape the target and be registered in a detector. If the new particles are LLPs, this gives them a better chance of escaping the target, although depending on their lifetime and boost, they may decay outside of the detector. Fixed-target experiments sensitive to LLPs include HPS \cite{Baltzell:2022rpd} (which is looking for a hidden version of electromagnetism mediated by a dark photon), NA62 \cite{NA62:2017rwk} (which studies rare kaon decays), and NA64 \cite{Andreas:2013lya} (which looks for invisible LLPs produced in the collision).

For relatively light LLPs with very long lifetimes, there is an intriguing class of experimental venue to search for them: beam dumps. Every accelerator must have a beam dump, which is the final destination of beam particles that do not undergo any collisions within the experiments. Unlike fixed targets, beam dumps are designed to be thick enough to absorb the entire beam. LLPs produced in the dump may fly out of it, and if they escape the beam dump entirely they can decay inside of a nearby detector. Examples of beam-dump experiments include E137 \cite{Bjorken:1988as}, which was a detector designed to look for particles produced in the SLAC beam dump, and NA62 \cite{NA62:2023qyn}, which is ordinarily a fixed target but can be turned into a beam dump to look for longer-lifetime LLPs.

There are a number of upcoming fixed-target and beam-dump experiments. These focus on BSM particle masses in the 10 MeV$/c^2$ to GeV$/c^2$ range:~this range is motivated by dark matter models that are not constrained by other experiments, as well as models of neutrino masses that can account for the matter-antimatter asymmetry. Proposed or upcoming experiments include SHiP \cite{SHiP:2015vad} and LDMX \cite{LDMX:2018cma}. 

\subsection{Astrophysics and Cosmology}
The Universe itself can be a laboratory for producing LLPs. There are many interesting connections between LLPs, astrophysics and cosmology, but we will only briefly touch on a few motivating examples.

All of the particles of the SM, and presumably most BSM particles too, were present in the hot and dense environment of the Big Bang. As the Universe expanded and cooled, many of these particles decayed away, and so today we are only left with the longest-lived particles, namely protons, electrons, neutrons, photons, and neutrinos. However, the evolution of the Universe could have been affected by LLP decays at earlier points. For example, the era of Big Bang Nucleosynthesis (BBN) took place when the Universe was about a second old: this is when the light elements formed out of proton and neutron building blocks. If an LLP decayed with a lifetime longer than about 1 s, however, the energetic decay products could have split nuclei apart and altered the physics of BBN. Since the standard BBN predictions agree extremely well with astrophysical data, LLPs should typically not have lifetimes longer than about 1 s.

LLPs can have other interesting astrophysical implications. For example, if dark matter interacts with SM particles, it can become trapped and accumulate inside of dense objects like stars and planets. Eventually, those dark matter particles annihilate, and if they annihilate to LLPs, those LLPs can escape the massive objects and decay to SM particles like photons or electrons in empty space. Those photons and electrons could then be picked up by cosmic ray telescopes, giving indirect evidence for dark matter. 

In fact, this phenomenon of weakly interacting LLPs produced in astrophysical objects has already been seen! In supernovae, photons become trapped inside of the dense core, but neutrinos (which feel only the weak force) escape from the supernova first. In 1987, a burst of neutrinos was seen from a supernova well before telescopes saw light emanating from the same event. In fact, supernova observations set strong constraints on new LLPs escaping supernovae which would lead to supernovae cooling in a different manner than we observe.

%%%%%%%%%%%%%%%%%%%%%%%%%%%%%%%%%%%%%%%%%
%% Mandatory: A concluding paragraph summing up your main points in the chapter
%% Optional: Also include big questions in the field that are still to be answered. What topics/methods/questions are researchers like to focus on next?
\section{Conclusions}
\label{sec:conclusions}
Long-lived particles are ubiquitous, both in the Standard Model and beyond. We have surveyed the dynamics responsible for long particle lifetimes, given examples of theories that include LLPs, and detailed the experimental opportunities and challenges in discovering new LLPs beyond the SM. We have focused on the Large Hadron Collider, but LLP searches are taking place at a wide range of energy scales in different types of experiments. LLP searches will be a cornerstone of any future program to discover new particles beyond the SM, and there is continuous development of proposals for innovative new LLP search strategies as well as refinements for current approaches.  \\

\begin{ack}[Acknowledgments]%
 We are grateful to the many collaborators and colleagues who have helped us over the years to develop the understanding of long-lived particles and their signatures which we present in this chapter. The authors thank Nathan Young for providing helpful comments on the draft.
\end{ack}

%%%%%%%%%%%%%%%%%%%%%%%%%%%%%%%%%%%%%%%%%
%% Mandatory: Bibliography using bibtex 
\bibliographystyle{Numbered-Style} %% for Numbered Reference Style
\bibliography{reference}

\end{document}